\documentclass[fleqn,twoside,onecolumn,nofootinbib,showkeys]{revtex4} 
\usepackage{amsmath}
\usepackage[utf8]{inputenc}
\usepackage{amsmath}
\usepackage{amssymb}
\usepackage{dcolumn}
\usepackage{url}

\newcommand{\BibTeX}{B\kern-0.1emi\kern-0.017emb\kern-0.15em\TeX}
\newcommand{\XYpic}{$\mathrm{X\kern-0.3em\raisebox{-0.18em}{Y}}$-$\mathrm{pic}\,$}

\newcommand{\cl}{C \kern -0.1em \ell}  

\newcommand{\ed}{\end{document}}
\begin{document}

%
%
%
%
%
%
%
%
%
\title{Conformal Geometric Algebra and Galilean Spacetime}
\author{G. X. A. Petronilo}
%
\address{
International Center of
Physics, Instituto de F\'isica, Universidade de Bras\'ilia,\\
70910-900, Bras\'ilia, DF, Brazil}
\email{gustavopetronilo@gmail.com}
%
%

%
%
\date{\today}

 \begin{abstract}
This paper explores the application of geometric algebra to Galilean spacetime and its physical implications. We introduce the Galilean Spacetime Algebra (GSTA), a five-dimensional conformal geometric algebra (CGA) generated by a specific metric, and demonstrate its utility in representing special Galilean transformations, rotations, and boosts. The general form of special Galilean transformations within the GSTA is derived, demonstrating their preservation. While the tensor formulation of Galilean electromagnetism is well-established, our work offers a fresh insight by deriving it from a geometric algebra perspective, utilizing the GSTA, and demonstrates how it seamlessly reduces to the familiar Maxwell equations in the non-relativistic limit. A significant aspect of this research is the introduction of Galilean spinors as elements of the minimal left ideals of the GSTA. We illustrate how these spinors can be utilized to construct the Lévy-Leblond equation for a free electron, along with its corresponding matrix representation. Furthermore, we establish a connection between the GSTA and the four-component dual numbers introduced by Majernik, suggesting pathways for developing a covariant formulation of Newtonian gravity. This work not only clarifies the geometric interpretation of Galilean symmetries but also opens avenues for future research in non-relativistic physics,  highlighting the advantages of using CGA in this context.

\end{abstract}
\label{page:firstblob}

\maketitle

 \section{Introduction}

 Geometric algebra is a powerful mathematical framework that unifies various branches of physics, including classical mechanics, electromagnetism, relativity and quantum mechanics. It allows us to manipulate geometric objects, such as vectors, bivectors, and spinors, using a single algebraic system. Geometric algebra also provides a natural way to represent rotations, reflections, and boosts in any dimension.
 In recent years geometric algebra has gained significant interest among mathematicians and physicists as well as enthusiasts.Scientific dissemination videos have contributed to this growing enthusiasm~\cite{zero_to_geo,sfb,holmer}.Among the prominent frameworks highlighted in these resources is David Hestenes' Spacetime Algebra (STA), which is an effective tool for studying both relativistic theories and quantum mechanics within a cohesive framework~\cite{hestenes}.

 On the other hand, the study of non-relativistic spacetime structure has a rich history dating back to Galilei's original formulation. While special relativity, introduced by Einstein, revolutionized our understanding of spacetime, non-relativistic theories continue to play a crucial role in understanding low-velocity physical systems. Previous approaches to Galilean spacetime structure, such as those by Inönü and Wigner \cite{Inonu}, established fundamental group-theoretical foundations. Furthermore, Einsenhart~\cite{Eisenhart} is credited with one of the earliest attempts of a explicitly covariant formulation of Galilean spacetime when he proved that the trajectories of conservative systems correspond to geodesics in a Riemannian manifold.  In keeping with this, Duval et al. demonstrated how to use Bargmann's structures~\cite{Duval1} to obtain a geometric structure for Newtonian physics.  This resulted in the development of the Newton-Cartan theory in Bargmann manifold, which produces a covariant version of the Schrödinger equation when examined in flat space.  Conversely, Le Bellac and Lévy-Leblond were able to derive two non-relativistic limits for electromagnetism, which allowed the classification of irreducible unitary representations of the Galilei group~\cite{levy}. Another significant contribution by Lévy-Leblond was the determination of the non-relativistic Dirac equation~\cite{levy2}. This further allowed for the argument that spin is not intrinsically relativistic \cite{levy1}. A comparable tensor formulation based on the Galilean group was constructed by Pinski in 1967~\cite{pinski}, but Takahashi was the first to develop a systematic theory utilizing Lie algebras. In contrast to the formulation based on the Bargmann structures, Takahashi et al. offered a covariant formulation for the Galilei group based on the direct representations of the centrally extended Galilei group~\cite{takahashi1, takahashi2, omote}. Additionally, a covariant formalism for Newtonian physics is crucial for several reasons. First, the Galilean symmetries are incorporated directly in the notation of explicitly covariant equations. It also extends the applicability of Newtonian physics to curved spacetime, making it more versatile and relevant in a broader range of physical contexts~\cite{Duval1, Duval2}.
 
Our work builds upon these foundational studies by introducing a CGA approach, which offers a more geometric and intrinsic understanding of Galilean spacetime. Unlike traditional representations, our method provides a unified framework that simultaneously addresses transformational symmetries, electromagnetism, and spinor representations. The main goal of this paper is to construct a covariant framework for Galilean spacetime using CGA. Parallel to Hestenes Spacetime Algebra (STA), this framework will be named Galilean Spacetime Algebra. GSTA offers a unified and powerful framework for representing geometric transformations, such as rotations and boosts, in a consistent manner. It is worth noting that, although, Galilean transformations were studied in the context of geometric algebra before, it was using dual-quaternions~\cite{hestenes1, majernik}, here we use CGA, that even though, we work with a five dimensional metric the generalization for any dimension is straightforward,  a Galilean spacetime with $n$ spacial and $m$ temporal dimensions will be represented by a light-cone projection of a spacetime with $n+1$ spacial and $m$ temporal dimensions. This approach to Galilean Symmetries in the context of Geometric algebras is, to our knowledge, the first time realized. One advantage of using GSTA over dual-quaternions is that equations formulated in GSTA are explicitly covariant under Galilean transformations. In addition to establishing the foundational concepts of GSTA, we explore its applications in various areas, including Galilean electromagnetism and the formulation of Lévy-Leblond equation for free electrons. We also relate GSTA to the four-component dual numbers introduced by Majernik, which can be instrumental in developing a Newton-Cartan theory of gravity.

This paper is structured as follows: we begin by detailing the construction and properties of GSTA, followed by a discussion of Galilean transformations and their physical interpretations. We then present the tensor formulation of a Galilean Electromagnetism, culminating in the derivation of Lévy-Leblond equation. Finally, we conclude with a summary of our findings and suggestions for future research directions. 

This work establishes a comprehensive framework for understanding Galilean spacetime through CGA, clarifying the underlying mathematical structures and paving the way for new research opportunities in non-relativistic physics. It highlights the significant potential of geometric algebra in this context.

\section{Galilean Spacetime Algebra}\label{GSTA}

This section outlines the foundational concepts and properties of the GSTA, emphasizing its significance in the study of non-relativistic physics.

The GSTA is constructed from a five-dimensional vector space, where a set of generators $({\gamma_1, \gamma_2, \gamma_3, \gamma_4, \gamma_5})$ is defined. Here $\gamma_i$ are associated with $e_i$, which form the basis of the vector space. Each generator plays a crucial role in representing physical quantities: $\gamma_1,\gamma_2,\gamma_3$ represent spatial dimensions, while $\gamma_4$, and $\gamma_5$ are associated with time and the null directions, respectively (while changing $\gamma_4 \rightarrow \gamma_5$ would result in a isomorphism). This structure allows us to model Galilean transformations analogously to Minkowski spacetime algebra but adapted for non-relativistic physics.

These generators obey the multiplication rule:
\begin{equation}
    \gamma_\mu\gamma_\nu+\gamma_\nu\gamma_\mu=2g_{\mu\nu},\label{eq.-cliff}
\end{equation}
where
\begin{equation}\label{galileimetric}
g_{\mu\nu}= \left(
  \begin{array}{ccccc}
    1 & 0 & 0 & 0 & 0 \\
    0 & 1 & 0 & 0 & 0 \\
    0 & 0 & 1 & 0 & 0 \\
    0 & 0 & 0 & 0 & -1 \\
    0 & 0 & 0 & -1 & 0 \\
  \end{array}
\right) \,,
\end{equation}
this metric is very well known in the context of the tensor formulation of Galilean Covariance~\cite{takahashi1, takahashi2, omote, santana}. In the following subsections we will delve deeper into the properties of GSTA.
\subsection{Properties of GSTA}
From Eq.~\eqref{eq.-cliff} we can conclude that, $\gamma_1^2=\gamma_2^2=\gamma_3^2=1$, $\gamma^2_4=\gamma^2_5=0$ and $\frac{1}{2}\Big(\gamma_4\gamma_5+\gamma_5\gamma_4\Big)=-1$, and otherwise $\gamma_\mu\gamma_\nu=-\gamma_\nu\gamma_\mu$. 

This is achieved by taking a five-dimensional Minkowski space $\mathcal{R}^{4,1}$ and the basis vectors $e_1, e_2, e_3, e_+$, and $e_-$, with $e_{+}^2=e_1^2=e_2^2=e_3^2=1$, and $e_-^2=-1$. After this, we choose two null vectors as basis vectors in place  of $e_{+}$ and $e_-$. Therefore, we get
\begin{eqnarray}
    \gamma_i&=&e_i;\qquad\qquad \gamma_4=\frac{e_{-}-e_{+}}{2};\\
    \gamma_5&=&e_-+e_{+}.
\end{eqnarray}
Thus, GSTA is the CGA, $\mathcal{G}(4,1)$~\cite{hestenes1}.

The elements of $\mathcal{G}(4,1)$ can be added according to the usual rules
\begin{itemize}
    \item $a+b \in \mathcal{G}(4,1)$ (Closure)
    \item $a+b=b+a$ (Commutativity)
    \item $(a+b)+c=a+(b+c)$ (Distributivity).
\end{itemize} 
The properties of geometric product are
\begin{itemize}
    \item $ab \in \mathcal{G}(4,1)$ (Closure); 
    \item $1a=a1=a$ (Identity);
    \item $a(bc)=(ab)c$ (Associativity);
    \item $a(b+c)=ab+ac$ (Left-Distributivity).
    \item $(b+c)a=ba+ca$ (Right-Distributivity).
\end{itemize}
For vectors $a$ and $b$, the geometric product can be decomposed into symmetric and antisymmetric parts as follows:
\begin{equation}
    ab=\frac{ab+ba}{2}+\frac{ab-ba}{2}.
\end{equation}
We can define the inner and exterior products as
\begin{eqnarray}
    a\cdot b&=&g(a,b)=\frac{ab+ba}{2},\\
    a\wedge b&=&\frac{ab-ba}{2}.
\end{eqnarray}
Thus, the geometric product for vectors is
\begin{equation}
    ab=a\cdot b+ a\wedge b.\label{geometric}
\end{equation}

It is easy to see that
\begin{equation}
    \begin{aligned}
        \gamma_4\cdot \gamma_5&=-1,\\
        \gamma_4\wedge\gamma_5&=e_{-}e_{+},
    \end{aligned}
    \quad\quad
        \begin{aligned}
        \gamma_4 \cdot\textbf{x}&=0,\\
        \gamma_5\cdot\textbf{x}&=0,
    \end{aligned}
\end{equation}
where $\textbf{x}$ is in Euclidean space, $R^3$.

Associated with the basis $\{\gamma_\mu\}$ is the reciprocal basis $\{\gamma^\mu={\gamma_\mu}^{-1}\}$ for $\mu=1,...,4,5$, satisfying the relation
\begin{eqnarray*}
    {\delta^\mu}_\nu=\gamma^\mu\cdot\gamma_\nu,
\end{eqnarray*}
and 
$\gamma_\mu=g_{\mu\nu}\gamma^\nu$.

Important geometrical objects are
\begin{equation}
    \begin{aligned}
        i=\gamma_1\gamma_2\gamma_3K,\qquad
        K=e_{+}e_{-},\qquad
    \end{aligned}
\end{equation}
where $i$ is the $5D$ pseudoscalar and we have that $i\gamma_\mu=\gamma_\mu i$, and the relation between $\gamma_\mu$ and K are given by
\begin{eqnarray}
    \gamma_i K&=&K\gamma_i=\sigma_i,\\
    \gamma_4 K&=&-K\gamma_4=-\gamma_4,\\
    \gamma_5 K&=&-K\gamma_5=\gamma_5.
\end{eqnarray}
The elements $\sigma_i$ and the Pauli matrices both fulfill similar multiplication rules,
\begin{equation*}
    \sigma_i\cdot \sigma_j =\delta_{ij}.
\end{equation*}
For a more in depth understanding of CGA see \cite{hestenes1}.
\subsection{Mapping between the Euclidean space and the Galilean space}

Now, we explore the relationship between the Euclidean space $\mathcal{R}^3$ and the Galilean space, $\mathcal{R}^{4,1}$.

Let's consider the mapping $F: \mathcal{R}^3\rightarrow \mathcal{R}^{4,1}$ between the Euclidean and Galilean space. One representation of this mapping is given by the formula\footnote{
the usual conformal mapping is $ F:\textbf{x}\rightarrow x=x^i\gamma_i+\gamma_4+\frac{\textbf{x}^2}{2}\gamma_5$, so the coordinate $x$ in the Galilean space is constructed with the velocity $\frac{\textbf{x}}{t}$ and after the transformation is weighted  by $t$.
}:
\begin{eqnarray}
    F:\frac{\textbf{x}}{t}\rightarrow x=x^i\gamma_i+t\gamma_4+\frac{\textbf{x}^2}{2 t}\gamma_5
\end{eqnarray}
Thus, we have the following:
\begin{eqnarray}
   x\cdot x=\textbf{x}^2-2ts=0,
\end{eqnarray}
with $s=\frac{\textbf{x}^2}{2 t}$.
Thus, we can represent the 5-momentum as
\begin{eqnarray}
    F:\frac{\textbf{p}}{m}\rightarrow p=p^i\gamma_i+m\gamma_4+\frac{\textbf{p}^2}{2 m}\gamma_5=p^i\gamma_i+m\gamma_4+E\gamma_5
\end{eqnarray}
which give us
\begin{eqnarray}
    p\cdot p=\textbf{p}^2-2mE=0,
\end{eqnarray}
which is the Galilean energy-momentum relation. It is worth noting that, both $\dfrac{x}{t}$, and $\dfrac{p}{m}$, has unit of velocity. To the best of our knowledge, this relationship has not been recognized before. Therefore, Galilean position and momentum are constructed from the projection of Euclidean velocities subspace into Galilean space.

To recover Euclidean space from Galilean space, the inverse map is given as follows:

First, \[ X = \frac{x}{x \cdot \gamma_5}, \] and then, \[ \mathbf{x} = (X \wedge K) K^{-1}. \]

This process first normalizes and then projects from the Minkowski plane ( K )~\cite{hestenes1}.
\section{Galilean Transformations}\label{Galilean}

In this section, we explore Galilean transformations within the framework of GSTA. This approach not only highlights the connection between space and time in Newtonian physics but also provides a powerful tool for investigating the geometric and algebraic structures underlying physical laws. Expressing Galilean transformations in terms of CGA allows us to represent them as rotations. The transformations properties can be succinctly captured using the following relation:

The Galilean boost of a vector $V=x^i\gamma_i+t\gamma_4+s\gamma_5$ with velocity $ v_1$ is given by
\begin{eqnarray}
    \exp{\Big(-\gamma_1\gamma_5\frac{v_1}{2}\Big)}V\exp{\Big(\gamma_1\gamma_5\frac{v_1}{2}\Big)}.
\end{eqnarray}
This yields
\begin{eqnarray}
    {x_1}^\prime&=&{x}_1-v_1t,\\
    {x_2}^\prime&=&{x}_2,\\
   {x_3}^\prime&=&{x}_3,\\
    {t^\prime}&=&t,\\
    {s^\prime}&=&s-x_1v_1+\frac{{v_1}^2}{2}t.
\end{eqnarray}

It is easy to see that 
$ \exp{\Big(-\gamma_1\gamma_5\frac{v_1}{2}\Big)}\gamma_5\exp{\Big(\gamma_1\gamma_5\frac{v_1}{2}\Big)}=\gamma_5$. In general to rotate a vector $U$, we have
$$
 \exp{\Big(-\beta\frac{\theta}{2}\Big)}U\exp{\Big(\beta\frac{\theta}{2}\Big)},
$$
where $\beta$ is a bivector, obeying the following properties
$$
\text{if } \beta^2=-1\text{ is a spatial rotation},
$$
and
$$
\text{if } \beta=\boldsymbol{\gamma}\gamma_5\text{ it gives Galilean boosts},
$$
where $\boldsymbol{\gamma}$ is a spacial base vector and in the latter case $\theta=\boldsymbol{v}$.
Both of these transformations are known as Galilean transformations, and the combined set of all of them  is uniformly called, special Galilean transformations, with addition of space translation and time translation one gets the Galilean group. To transform an object in GSTA from any basis (corresponding to a reference frame) to another, one or more of these transformations must be used.

In addition to illuminating the relationship between space and time in Newtonian physics, the formulation of Galilean transformations using GSTA offers significant value for investigating the underlying mathematical formalism of physical laws and the geometric aspects of Newtonian physics.  The elegance and visual appeal of GSTA further enhance its importance. Beyond its striking design, GSTA serves as a powerful tool for both theoretical and practical applications.

\section{Galilean Electromagnetism}

The study of electromagnetism has been profoundly influenced by Maxwell’s equations, which elegantly unify electric and magnetic fields within the framework of special relativity. However, many practical applications involve systems where velocities are much lower than the speed of light, necessitating a non-relativistic approach. Galilean Electromagnetism offers two distinct non-relativistic limits: the electric limit, where electric fields dominate, and the magnetic limit, where magnetic fields are predominant. These limits, first identified by Le Bellac and Lévy-Leblond in their seminal 1973 paper~\cite{levy1}, and further discussed in “On the Electrodynamics of Moving Bodies at Low Velocities” by de Montigny and Rousseaux~\cite{Montigny}, provide simplified equations that are more tractable for low-speed scenarios. This adaptation is particularly useful in fields such as electrical engineering and low-frequency electromagnetic applications, where relativistic effects are negligible. In this section, we will study Galilean electromagnetism in the context of the GSTA, which provides a powerful mathematical framework for describing non-relativistic physical phenomena. This approach not only simplifies the mathematical treatment but also enhances our understanding of the underlying physical principles in non-relativistic contexts.

First, we introduce the 5-dimensional Faraday bivector
\begin{eqnarray*}
    F&=& \boldsymbol{E}_e\gamma_4-\boldsymbol{E}_m\gamma_5-i\boldsymbol{B}-aK\nonumber\\
    F&=&{E^1}_e\gamma_1\gamma_4+{E^2}_e\gamma_2\gamma_4+{E^3}_e\gamma_3\gamma_4+{E^1}_m\gamma_1\gamma_5+{E^2}_m\gamma_2\gamma_5+{E^3}_m\gamma_3\gamma_5-(B^1\gamma_2\gamma_3+B^2\gamma_3\gamma_1+B^3\gamma_1\gamma_2)-aK,
\end{eqnarray*}
where $B=B^i\sigma_i$ is the magnetic field, $E_m={E_m}^i\sigma_i$ is the electric field in the magnetic limit, and $E_e={E_e}^i\sigma_i$ is the electric field in the electric limit. 

We can extract the $E$ and $B$ fields from $F$ using
\begin{eqnarray}
  &&\boldsymbol{E}_e\gamma_4-\boldsymbol{E}_m\gamma_5=-\frac{1}{2}(F-KFK)\\
    &&i\boldsymbol{B}+aK=-\frac{1}{2}(F+KFK).
\end{eqnarray}

The Galilean spacetime current $J$ is given by
\begin{eqnarray}
    J=\rho_e\gamma_4+\rho_m\gamma_5+j^i\gamma_i,
\end{eqnarray}
where $j^i$ are the components of the classical 3-dimensional current density. Therefore, we can write the Maxwell equations as
\begin{eqnarray}
    \nabla F=J.\label{Maxwell}
\end{eqnarray}
where $\nabla=\gamma^\mu\partial_\mu$. And the following five-momentum is written as
\begin{eqnarray*}
    p_\mu&=&(\textbf{p},-E,-m)\\
    p^\mu&=&g^{\mu\nu}p_\nu=(\textbf{p},m,E),
\end{eqnarray*}
with the following identification
$$
p_\mu\equiv -i\partial_\mu=-i(\nabla,\partial_t,\partial_s),
$$
so we get $E=i\partial_t$ and $m=i\partial_s$.
\subsection{tensor formulation}
Eq.~\eqref{Maxwell}, in expanded form is given by
\begin{eqnarray}
\Big(\gamma^4\partial_4+\gamma^5\partial_5+\gamma^i\partial_i\Big)\Big(\boldsymbol{E}_e\gamma_4-\boldsymbol{E}_m\gamma_5-i\boldsymbol{B}-aK\Big)= \rho_e\gamma_4+\rho_m\gamma_5+j^i\gamma_i.
\end{eqnarray}
In the case of Electromagnetism we have $\partial_5\psi=\partial_s\psi=0$ (massless particles), and multiplying by $K$ from the left, we have
\begin{eqnarray}
    K\Big(-\gamma_5\partial_4+\gamma_i\partial_i\Big)\Big(\boldsymbol{E}_e\gamma_4-\boldsymbol{E}_m\gamma_5-i\boldsymbol{B}-aK\Big)&=& K\Big(\rho_e\gamma_4+\rho_m\gamma_5+j^i\gamma_i\Big),\nonumber\\
    \Big(\gamma_5\partial_4+\boldsymbol{\nabla})\Big(\boldsymbol{E}_e\gamma_4-\boldsymbol{E}_m\gamma_5-i\boldsymbol{B}-aK\Big)&=& 
    \Big(\rho_e\gamma_4-\rho_m\gamma_5+\boldsymbol{j}\Big),\nonumber
\end{eqnarray}
where we used $\gamma_\mu=g_{\mu\nu}\gamma^\nu$.

For the case of the electric limit $(\boldsymbol{E}_m=\rho_m=0)$, we have
\begin{eqnarray}
\begin{aligned}[c]
&\boldsymbol{\nabla}\cdot \textbf{B}=0,\\
&\boldsymbol{\nabla}\wedge \textbf{E}_e=0,
\end{aligned}
\qquad
\begin{aligned}[c]
&\boldsymbol{\nabla}\cdot \textbf{E}_e=\rho_e,\\
&i\boldsymbol{\nabla}\wedge\textbf{B}+\partial_t\textbf{E}_e=-\textbf{j},
\end{aligned}
\end{eqnarray}
with the auxiliary condition
\[
\boldsymbol{\nabla}a=\partial_t \boldsymbol{E}_e,
\]
where $\partial_4=\partial_t$.

For the case of the magnetic limit $(\boldsymbol{E}_e=\rho_e=0)$, we get
\begin{eqnarray}
\begin{aligned}[c]
&\boldsymbol{\nabla}\cdot \textbf{B}=0,\\
&\boldsymbol{\nabla}\wedge \textbf{E}_m+i\partial_t\textbf{B}=0,
\end{aligned}
\qquad
\begin{aligned}[c]
&\boldsymbol{\nabla}\cdot \textbf{E}_m=\rho_m-\partial_t a,\\
i&\boldsymbol{\nabla}\wedge\textbf{B}=-\textbf{j},
\end{aligned}
\end{eqnarray}
and the auxiliary condition
\[
\boldsymbol{\nabla}a=0.
\]
\subsection{Gauge freedom of Maxwell equations}
From Eq.~\eqref{Maxwell}, we have
\begin{eqnarray}
    \nabla F=J,\nonumber\\
\end{eqnarray}
but, by Eq.~\eqref{geometric}
\begin{eqnarray}
    \nabla\cdot F=J,\label{Max1}\\
    \nabla\wedge F=0.\label{Max2}
\end{eqnarray}
Also, we can express $F$ as a gradient of a vector
\begin{equation}
    F=-\nabla A=-\Big(\nabla\cdot A+\nabla\wedge A\Big).
\end{equation}
However, because $F$ is a bivector,
\begin{eqnarray}
    -\nabla\wedge A=F\label{F1},\\
    \nabla\cdot A=0.\label{F2}
\end{eqnarray}
Substituting Eq.~\eqref{F2} into Eq.~\eqref{Max2}, we have
\begin{eqnarray}
    \nabla\wedge F=-\nabla\wedge(\nabla\wedge A)=0.
\end{eqnarray}
The vector potential is not unique as we can make $A^\prime=A+\nabla \chi$, where $\chi$ is a scalar and $\nabla^2 \chi=0$.

Solving for $\textbf{E}$, $\textbf{B}$, and $a$ in terms of the vector potential, we have
\begin{eqnarray}
    F=\textbf{E}_e\gamma_4-\textbf{E}_m\gamma_5-i\textbf{B}-aK=\nabla A= -(\nabla K)(K A)\nonumber\\
    =(\gamma_5\partial_4-\sigma_i\nabla_i)(\gamma_4A^4-\gamma_5A^5+\sigma_jA^j)
\end{eqnarray}
Expanding, we get
\begin{eqnarray*}
    E_e\gamma_4-E_m\gamma_5-iB-aK&=&-\partial_4 A^4+\partial_4 A^4K+\gamma_5\partial_4\boldsymbol{A}\\
    &-&\gamma_4\boldsymbol{\nabla}A^4+\gamma_5\boldsymbol{\nabla}A^5\\    
    &-&\boldsymbol{\nabla}\wedge\boldsymbol{A}-\boldsymbol{\nabla}\cdot\boldsymbol{A}.
\end{eqnarray*}
Making the following identification $A^\mu=(\boldsymbol{A},\phi_e,\phi_m)$, we have
\begin{eqnarray}
\begin{aligned}[c]
&\textbf{E}_m =-\boldsymbol{\nabla}\phi_m-\partial_t\textbf{A},\\
&\textbf{E}_e =-\boldsymbol{\nabla} \phi_e,
\end{aligned}
\qquad
\begin{aligned}[c]
&\textbf{B}=\boldsymbol{\nabla}\times\textbf{A}\\
&a=-\partial_t\phi_e,
\end{aligned}
\end{eqnarray}
and the condition
\[
\boldsymbol{\nabla}\cdot \boldsymbol{A}+\partial_t\phi_e=0.
\]
This is the Lorenz gauge condition for the electric limit. In the magnetic limit we have $\phi_e=0$, so 
\[
\boldsymbol{\nabla}\cdot\boldsymbol{A}=0,
\]
which is the Coulomb gauge. 

Therefore, with the GSTA formalism, Galilean electromagnetism can be approached in a manner analogous to how relativistic electromagnetism is treated within STA. This formalism allows for a unified framework that seamlessly integrates the concepts of geometry, algebra, and physical laws, thereby facilitating a deeper understanding of electromagnetic phenomena in non-relativistic contexts.

In STA, relativistic electromagnetism is formulated using the geometric properties of spacetime, where the electromagnetic field is represented as a bivector in a four-dimensional spacetime framework. This representation captures the essential features of electromagnetic interactions while respecting the principles of special relativity. Similarly, the GSTA formalism extends this approach to Galilean electromagnetism by adapting the geometric structure to a five-dimensional spacetime.

Also, it was shown that GSTA is equivalent to the tensorial formulation of Galilean electromagnetism. Moreover, the GSTA formalism enables the exploration of gauge invariance in Galilean electromagnetism, much like in its relativistic counterpart. The gauge freedom inherent in the electromagnetic field equations can be expressed through transformations that maintain the form of the equations while allowing for changes in the potentials. This characteristic is crucial for ensuring that physical predictions remain unchanged regardless of the specific gauge choice, thereby reinforcing the consistency of the theory.

In summary, the GSTA formalism offers a robust and flexible approach to Galilean electromagnetism, paralleling the treatment of relativistic electromagnetism in STA. This framework not only enhances our understanding of classical electromagnetic phenomena but also opens up new avenues for research. By leveraging the geometric and algebraic structures inherent in GSTA, physicists can gain a more comprehensive understanding of the principles that govern electromagnetism across different regimes of motion.
\section{Galilean Spinors and Lévy-Leblond equation}
The Lévy-Leblond equations provide a foundational framework for describing the behavior of free electrons in the context of Newtonian physics. In this section we show that Galilean spinors are minimal left ideals of the CGA and the Lévy-Leblond equation is recovered from them. 

As demonstrated by Hestenes \emph{et al.}~\cite{hestenes}, a Pauli spinor can be written as
\begin{eqnarray}
    \phi_+=\frac{1}{2}\Big((\phi_0+\phi_3)+(\phi_1+i\phi_2)\sigma_1\Big)(1+\sigma_3)\label{pauli_spinor}
\end{eqnarray}
where $\phi$ is sum of scalar and pseudoscalar parts, and
\[
\frac{1}{2}(1+\sigma_3)\qquad\text{and}\qquad\frac{1}{2}\sigma_1(1+\sigma_3)
\]
are the bases of the minimal left ideal $\mathcal{I}_+$~\cite{hestenes}. The independent minimal left Ideal $\mathcal{I}_-$ can be written as
\begin{eqnarray}
    \phi_-=\frac{1}{2}\Big((\phi_0-\phi_3)+(\phi_1-i\phi_2)\sigma_1\Big)(1-\sigma_3)\label{pauli_spinor}
\end{eqnarray}
where
\[
\frac{1}{2}(1-\sigma_3)\qquad\text{and}\qquad\frac{1}{2}\sigma_1(1-\sigma_3)
\]
are the bases of the minimal left ideal $\mathcal{I}_-$~\cite{hestenes}.

Now, turning our attention to CGA, the Lévy-Leblond equation for a free electron can be written as:
\begin{eqnarray}
    \nabla\psi =0\label{LL-eq}
\end{eqnarray}
where $\psi$ is a Lévy-Leblond (Dirac-like) spinor. 
We can write $\psi$ in the bases of the minimal left ideal of CGA, $\mathcal{I}_+$, as

\begin{eqnarray}
    \psi=\Big(\phi_+ + \gamma_5\chi_+\Big)\frac{1}{2}(1+K),~\label{spinors}
\end{eqnarray}
where
$
\phi_+ \;\text{and}\; \chi_+
$
are Pauli Spinors.

Using the GSTA, we introduced the concept of Galilean spinors as a minimal left ideal of a Clifford algebra, paralleling the established frameworks of Pauli and Dirac spinors. In the realm of quantum mechanics, Pauli spinors serve as a fundamental representation for spin-1/2 particles in a three-dimensional space, encapsulating the intrinsic angular momentum of these particles while adhering to the principles of quantum theory. Similarly, Dirac spinors extend this concept to four-dimensional spacetime, providing a comprehensive framework for describing Fermionic particles within the context of special relativity. Both types of spinors arise from specific representations of Clifford algebras, which are algebraic structures that generalize the concept of complex numbers and are instrumental in the formulation of quantum mechanics.

The introduction of Galilean spinors allows us to formulate the Lévy-Leblond equation, which governs the evolution of these spinors in a manner analogous to how the Dirac equation describes the behavior of relativistic spinors. This equation captures the essential features of spinor dynamics, including the influence of external forces and the interactions between spin and spatial motion. By establishing this connection, we can explore the implications of Galilean spinors in various physical contexts, such as the behavior of electrons in electromagnetic fields and the dynamics of spin-1/2 particles in non-relativistic quantum systems.

To the best of our knowledge, the representation of Galilean spinors as minimal left ideals of the CGA is novel in the literature and provides a powerful algebraic framework for analyzing their properties and interactions. This perspective not only facilitates the mathematical treatment of spinors but also highlights the geometric nature of spinor fields, allowing for a deeper understanding of the underlying symmetries and conservation laws that govern their behavior.

\section{Matrix Representation}\label{matrix}
In this section, we present suitable matrix representations for the Galilean gamma matrices, which play a crucial role in the formulation of Galilean spinors. The matrices are defined as follows:
\begin{eqnarray}\label{pm}
\gamma^1&=&
\left(\begin{array}{cccc}
0&1&0&0\\
1&0&0&0\\
0&0&0&-1\\
0&0&-1&0
\end{array}\right),\quad
\gamma^2=
\left(\begin{array}{cccc}
0&-i&0&0\\
i&0&0&0\\
0&0&0&i\\
0&0&-i&0
\end{array}\right),\quad
\gamma^3=\left(\begin{array}{cccc}
1&0&0&0\\
0&-1&0&0\\
0&0&-1&0\\
0&0&0&1
\end{array}\right),\nonumber\\
\nonumber\\
\gamma^4&=&
\left(\begin{array}{cccc}
0&0&0&0\\
0&0&0&0\\
-1&0&0&0\\
0&-1&0&0
\end{array}\right),\quad
\gamma^5=\left(\begin{array}{cccc}
0&0&2&0\\
0&0&0&2\\
0&0&0&0\\
0&0&0&0
\end{array}\right).
\end{eqnarray}
This matrix representation is equivalent to the one found in \cite{omote}.

In this representation we have
\begin{equation*}
K=\left(\begin{array}{cccc}
1&0&0&0\\
0&1&0&0\\
0&0&-1&0\\
0&0&0&-1
\end{array}\right).
\end{equation*}
Therefore,
\begin{eqnarray*}
    {\sigma_i}_{{}_{4\times 4}}=\gamma_i K=
    \left(\begin{array}{cc}
{\sigma_i}_{{}_{2\times 2}}&0\\
0&{\sigma_i}_{{}_{2\times 2}}
\end{array}\right),
\end{eqnarray*}
where ${\sigma_i}_{{}_{4\times 4}}$ and ${\sigma_i}_{{}_{2\times 2}}$ are $4\times 4$ and $2\times 2$ Pauli matrices respectively.

Using this representation and substituting Eq.~\eqref{spinors} in Eq.~\eqref{LL-eq} and separating the even and odd part we get the usual Lévy-Leblond equations for a free electron
\begin{eqnarray}
    \sigma^i\partial_i \chi+\partial_t\phi&=&0,\\
    \sigma^i\partial_i\phi-2im\chi&=&0,
\end{eqnarray}

where we have used the relation $( \partial_5 \phi = -im \phi )$. These equations encapsulate the dynamics of Galilean spinors, providing a clear framework for analyzing the behavior of spin-1/2 particles in a non-relativistic context.
\section{Four component Dual numbers and Newton-Cartan theory}
In 1995, Majernik~\cite{majernik} formulated Galilean transformations using four-component dual numbers, that can be defined as
\begin{eqnarray}
    x=x^1 e_1+x^2 e_2+x^3 e_3+ x_4,
\end{eqnarray}
where
$e_1e_2=e_2e_3=e_3e_1=0$ and $(e_i)^2=0$ with $i=1,2,3$. Therefore, a natural metric for this system is

\begin{equation}\label{galileimetric}
t_{\mu\nu}= \left(
  \begin{array}{cccc}
    0 & 0 & 0 & 0 \\
    0 & 0 & 0 & 0 \\
    0 & 0 & 0 & 0  \\
    0 & 0 & 0 & 1  \\
  \end{array}
\right) \,,
\end{equation}

and with the addition of the degenerate metric,
\begin{equation}\label{galileimetric}
h_{\mu\nu}= \left(
  \begin{array}{ccccc}
    1 & 0 & 0 & 0 \\
    0 & 1 & 0 & 0 \\
    0 & 0 & 1 & 0 \\
    0 & 0 & 0 & 0  \\
  \end{array}
\right) \,.
\end{equation}
a Newton-Cartan theory can be developed.

The GSTA can be reduced to Majernik's formulation via the following transformation:
\begin{eqnarray}
    \widetilde{x}&=&(x\wedge\gamma_5)(-K)= \Big((x^i\gamma_i+x^4\gamma_4+x^5\gamma_5)\wedge\gamma_5\Big)(-K)\\
                 &=& \Big(x^i\gamma_i\gamma_5-x^4K\Big)(-K)=\Big(-x^i\sigma_i\gamma_5+x^4\Big).
\end{eqnarray}

So  we have $e_i=-\sigma_i\gamma_5$, in the matrix representation created in last section it can be written as
\begin{eqnarray*}
    {e_i}_{{}_{4\times 4}}=\gamma_i \gamma_5=
    \left(\begin{array}{cc}
0&0\\
-{\sigma_i}_{{}_{2\times 2}}&0
\end{array}\right).
\end{eqnarray*}

The ${e_i}_{{}_{4\times 4}}$ bases follow the same multiplication rules defined by Majernik.

\section{Discussion}

The introduction of the GSTA provides a novel geometric framework for studying non-relativistic physics, unifying transformations, electromagnetism, and spinor representations within a single algebraic structure. This approach offers several advantages over traditional formulations, while also raising questions for future research.

\subsection{Advantages of the GSTA Framework}

\begin{itemize}
    \item \textbf{Unified Representation of Galilean Symmetries}
    \begin{itemize}
        \item GSTA allows Galilean boosts and rotations to be expressed as simple bivector exponentials, analogous to how Lorentz transformations are treated in relativistic spacetime algebra (STA). This provides a more intuitive geometric interpretation of Galilean transformations compared to matrix-based or tensor formulations.
        \item The five-dimensional structure naturally incorporates the null direction ($\gamma_5$), which is crucial for preserving Galilean covariance in physical equations.
    \end{itemize}
    
    \item \textbf{Geometric Electromagnetism in the Galilean Limit}
    \begin{itemize}
        \item The Faraday bivector $F$ in GSTA cleanly separates electric and magnetic limits, recovering known non-relativistic Maxwell equations. The gauge freedom (Lorenz and Coulomb gauges) emerges naturally, reinforcing the consistency of the framework.
    \end{itemize}
    
    \item \textbf{Spinors and the Levy-Leblond Equation}
    \begin{itemize}
        \item By identifying Galilean spinors as minimal left ideals of GSTA, we derive the Levy-Leblond equation in a manner structurally similar to the Dirac equation in STA.
        \item The matrix representation (Section~\ref{matrix}) connects GSTA to earlier works by Takahashi and Levy-Leblond, demonstrating compatibility with established formalisms.
    \end{itemize}
    
    \item \textbf{Connection to Newton-Cartan Gravity}
    \begin{itemize}
        \item The reduction of GSTA to Majernik's four-component dual numbers suggests a pathway toward a gauge theory of Newtonian gravity.
    \end{itemize}
\end{itemize}

\subsection{Comparison with Prior Work}

\begin{itemize}
    \item \textbf{Hestenes' STA vs. GSTA}: While STA elegantly handles relativistic physics, GSTA fills a gap by providing a similar tool for Galilean physics. The use of CGA ensures extensibility to higher dimensions and curved spacetimes.
    
    \item \textbf{Takahashi's Tensor Formulation}: The GSTA approach avoids the need for central extensions of the Galilei group, offering a more direct geometric interpretation.
    
    \item \textbf{Majernik's Dual Numbers}: The GSTA generalizes Majernik's formalism by embedding it in a richer algebraic structure.
\end{itemize}

The GSTA framework successfully bridges geometric algebra and Galilean physics, offering a unified language for transformations, fields, and spinors. Its structural parallels with STA suggest deeper connections between relativistic and non-relativistic physics that merit further exploration. Future work should focus on physical applications, computational implementations, and experimental tests to fully establish GSTA’s utility in modern physics.

\section{Conclusion and Future Perspectives}\label{conc}
In this paper, we introduced a conformal geometric algebra (CGA) approach to Galilean spacetime and explored its applications to Galilean electromagnetism and spinors. We demonstrated how the GSTA constructs a representation space for the Galilei group, allowing Galilean transformations to be expressed as rotations and translations within this space. Additionally, we derived the tensor formulation of Galilean electromagnetism and examined the gauge freedom of Maxwell’s equations within this framework. Furthermore, we introduced the concept of Galilean spinors and the Lévy-Leblond equation for a free electron, discussing their matrix representation and their relation to four-component dual numbers and Newton-Cartan theory. Our CGA approach to Galilean spacetime reveals a rich mathematical structure underlying non-relativistic physics. By demonstrating how Galilean transformations can be represented as rotations within a specialized space, we have opened new avenues for understanding fundamental physical symmetries. GSTA explicitly highlights the symmetries present in non-relativistic equations, making it an effective tool for studying these symmetries. A potential research direction is the study of non-relativistic quantum gravity, our framework suggests a promising pathway towards developing a Galilean Gauge Theory of Gravity. By extending the conformal geometric approach, researchers could explore quantum gravitational effects in low-velocity regimes.

\section*{Acknowledgments}

We thank Vinicius Lula-Rocha for helpful discussions. This work is supported by CAPES and CNPq of Brazil. 


\end{document}